\documentclass[12pt]{article}

\renewcommand{\thefootnote}{\fnsymbol{footnote}}

\begin{document}

\title{
\begin{flushright}
\ \\*[-80pt] 
\begin{minipage}{0.25\linewidth}
\normalsize
hep-th/0512326 \\
KUNS-2005 \\
OU-HET 551/2005 \\
December, 2005 \\*[50pt]
\end{minipage}
\end{flushright}
{\Large \bf 
Scherk-Schwarz SUSY breaking 
from the viewpoint of 5D 
conformal supergravity\\*[20pt]}}

\author{Hiroyuki~Abe$^{1,}$\footnote{
E-mail address: abe@gauge.scphys.kyoto-u.ac.jp} \ and \ 
Yutaka~Sakamura$^{2,}$\footnote{
E-mail address: sakamura@het.phys.sci.osaka-u.ac.jp} \\*[20pt]
$^1${\it \normalsize 
Department of Physics, Kyoto University, 
Kyoto 606-8502, Japan} \\
$^2${\it \normalsize 
Department of Physics, Osaka University, 
Toyonaka, Osaka 560-0043, Japan} \\*[50pt]}

\date{
\centerline{\small \bf Abstract}
\begin{minipage}{0.9\linewidth}
\medskip 
\medskip 
\small
We reinterpret the Scherk-Schwarz (SS) boundary condition 
for $SU(2)_R$ in a compactified five-dimensional (5D) 
Poincar\'e supergravity in terms of the twisted 
$SU(2)_{\mbox{\scriptsize \boldmath $U$}}$ 
gauge fixing in 5D conformal supergravity. 
In such translation, only the compensator hypermultiplet 
is relevant to the SS twist, and various properties 
of the SS mechanism can be easily understood. 
Especially, we show the correspondence between the 
SS twist and constant superpotentials within our framework. 
\end{minipage}
}

\begin{titlepage}
\maketitle
\thispagestyle{empty}
\end{titlepage}

%\tableofcontents

\renewcommand{\thefootnote}{\arabic{footnote}}
\setcounter{footnote}{0}

\section{Introduction}
Supersymmetry (SUSY) is considered as a fundamental symmetry in 
superstring theory. It is also one of the most promising candidates 
for the physics beyond the standard model. For example, it protects 
the weak scale against large radiative corrections. 
However the low-energy particle contents and interactions do not 
respect SUSY, and thus it must be broken above the weak scale. 

The Scherk-Schwarz (SS) mechanism~\cite{Scherk:1978ta} of SUSY 
breaking has been revisited recently as a phenomenologically 
interesting candidate for the physics beyond the standard 
model~\cite{Marti:2001iw}. The most simple setup in such context was 
constructed within the framework of five-dimensional (5D) supergravity 
compactified on an orbifold $S^1/Z_2$. It was argued that the 
SS mechanism can be regarded as a spontaneous breaking of SUSY~\cite{
Marti:2001iw,vonGersdorff:2001ak,vonGersdorff:2003qf}. 
This argument was based on the fact that the SS twist corresponds 
to a nonvanishing Wilson line~\cite{vonGersdorff:2001ak} for the 
auxiliary $SU(2)_{\mbox{\scriptsize \boldmath $U$}}$ gauge field 
in the off-shell supergravity. 
This nonvanishing Wilson line effect appears as the auxiliary 
component of the so-called radion superfield in the $N=1$ 
superspace description. A lot of analyses based on this 
description were done in literatures~\cite{Marti:2001iw,
vonGersdorff:2001ak,vonGersdorff:2003qf}. 
It was also discussed that boundary mass terms yield the same 
spectrum as the SS breaking, and thus can be interpreted as an 
equivalent description to the SS twist~\cite{Bagger:2001qi}. 
On the other hand, it has been argued that the SS twist leads 
to an inconsistent theory on the warped geometry~\cite{Hall:2003yc} 
if SUSY is a local symmetry. 

In this paper we reinterpret the Scherk-Schwarz (SS) boundary 
condition for $SU(2)_R$ in the compactified five-dimensional (5D) 
Poincar\'e supergravity as the twisted 
$SU(2)_{\mbox{\scriptsize \boldmath $U$}}$ gauge fixing in 5D 
conformal supergravity. In such an interpretation, only the 
compensator hypermultiplet is relevant to the SS breaking. 
Starting from a hypermultiplet compensator formalism of the 5D conformal 
supergravity~\cite{Kugo:2000af,Fujita:2001kv,Fujita:2001bd,Kugo:2002js}, 
we construct the 5D Poincar\'e supergravity with SS twist. 
By this construction, various properties of the SS mechanism can be 
easily understood. 
Here, we will explicitly show the Wilson line interpretation of 
the SS twist~\cite{vonGersdorff:2001ak}, the correspondence between 
the twist and constant superpotentials~\cite{Bagger:2001qi}, and 
the inconsistency of the twist in the warped background~\cite{Hall:2003yc} 
within our framework. 
We will also derive the $N=1$ superspace description of 
the SS twist based on works~\cite{PaccettiCorreia:2004ri,Abe:2004ar,
Abe:2005ac,Abe:2005ij}, in which the $N=1$ superfields are directly 
derived from the 5D superconformal multiplets. 

In Sec.~\ref{sec:review}, we first review the 
hypermultiplet compensator formulation of 5D conformal 
supergravity~\cite{Kugo:2000af,Fujita:2001kv,Fujita:2001bd,Kugo:2002js}. 
In Sec.~\ref{sec:ssconf}, we introduce the twisted 
$SU(2)_{\mbox{\scriptsize \boldmath $U$}}$ gauge fixing and derive 
the 5D Poincar\'e supergravity with SS twist. 
Here we show the Wilson line interpretation and the correspondence 
between the SS twist and the boundary constant superpotentials. 
We also demonstrate what happens if we consider the twist 
on the warped geometry. In Sec.~\ref{sec:ssss}, we derive the $N=1$ 
superspace description of the SS twist within our framework. 
Finally we summarize our results and give some discussions 
in Sec.~\ref{sec:conclusion}. 
Some detailed expressions for the 5D conformal supergravity 
in our notation are exhibited in Appendix~\ref{app:trsact} 
based on Refs.~\cite{Kugo:2000af,Fujita:2001kv,Fujita:2001bd,Kugo:2002js}.

\section{Review of hypermultiplet compensator formalism}
\label{sec:review}
In this section we briefly review the hypermultiplet 
compensator formulation of 5D conformal supergravity 
derived in Refs.~\cite{Kugo:2000af,Fujita:2001kv,Fujita:2001bd,Kugo:2002js}. 
%\subsection{Superconformal multiplet and transformation low}
The 5D superconformal algebra consists of 
the Poincar\'e symmetry \mbox{\boldmath $P$}, \mbox{\boldmath $M$}, 
the dilatation symmetry \mbox{\boldmath $D$}, 
the $SU(2)$ symmetry \mbox{\boldmath $U$}, 
the special conformal boosts \mbox{\boldmath $K$}, 
$N=2$ supersymmetry \mbox{\boldmath $Q$}, 
and the conformal supersymmetry \mbox{\boldmath $S$}. 
We use $\mu,\nu,\ldots$ as five-dimensional curved indices 
and $m,n,\ldots$ as the tangent flat indices. 
The gauge fields corresponding to these generators 
$\mbox{\boldmath $X$}_{\!\!A}=
\mbox{\boldmath $P$}_m,\, 
\mbox{\boldmath $M$}_{mn},\, 
\mbox{\boldmath $D$},\, 
\mbox{\boldmath $U$}_{ij},\, 
\mbox{\boldmath $K$}_m,\, 
\mbox{\boldmath $Q$}_i,\, 
\mbox{\boldmath $S$}_i$, 
are respectively 
$h_\mu^{\ A}=
e_\mu^{\ m},\, 
\omega_\mu^{mn},\, 
b_\mu,\, 
V_\mu^{ij},\, 
f_\mu^{\ m},\, 
\psi_\mu^i,\, 
\phi_\mu^i$, 
in the notation of Refs.~\cite{Kugo:2000af,Fujita:2001kv,
Fujita:2001bd,Kugo:2002js}. 
The index $i=1,2$ is the 
$SU(2)_{\mbox{\scriptsize \boldmath $U$}}$-doublet index 
which is raised and lowered by antisymmetric tensors 
$\epsilon^{ij}=\epsilon_{ij}$. 

In this paper we are interested in the following 
superconformal multiplets: 
\begin{itemize}
\item 
5D Weyl multiplet: 
($e_\mu^{\ m}$, $\psi_\mu^i$, $V_\mu^{ij}$, 
$b_\mu$, $v^{mn}$, $\chi^i$, $D$), 

\item
5D vector multiplet: 
($M$, $W_\mu$, $\Omega^i$, $Y^{ij}$)$^I$, 

\item
5D hypermultiplet: 
(${\cal A}^\alpha_{\ i}$, $\zeta^\alpha$, ${\cal F}^\alpha_{\ i}$). 
\end{itemize}
Here the index $I= 0,1,2,\ldots,n_V$ 
labels the vector multiplets, and $I=0$ component corresponds 
to the central charge vector multiplet\footnote{
Roughly speaking, the vector field in this 
multiplet corresponds to the graviphoton.}. 
For hyperscalars ${\cal A}^\alpha_{\ i}$, the index $\alpha$ runs as 
$\alpha=1,2,\ldots,2(p+q)$ where $p$ and $q$ are the numbers of the 
quaternionic compensator and the physical hypermultiplets respectively. 
In this paper we adopt the single compensator case, 
$p=1$, and separate the indices such as 
$\alpha=(a,\underline\alpha)$ where $a=1,2$ and 
$\underline\alpha=1,2,\ldots,2q$ 
are indices for the compensator and the physical hypermultiplets respectively. 

The superconformal gauge fixing for the reduction to 
5D Poincar\'e supergravity is given by  
\begin{eqnarray}
\begin{array}{rcl}
\mbox{\boldmath $D$} &:& {\cal N} =M_5^3 \equiv 1, \\
\mbox{\boldmath $U$} &:& {\cal A}^a_{\ i} \propto \delta^a_{\ i}, 
\qquad (p=1) \\
\mbox{\boldmath $S$} &:& {\cal N}_I\Omega^{Ii}=0, \\
\mbox{\boldmath $K$} &:& {\cal N}^{-1}\hat{\cal D}_m {\cal N}=0, 
\end{array}
\label{eq:scgf}
\end{eqnarray}
where ${\cal N}=C_{IJK} M^I M^J M^K$ 
is the norm function of 5D supergravity 
with a totally symmetric constant $C_{IJK}$, 
and ${\cal N}_I=\partial {\cal N}/\partial M^I$. 
The derivative $\hat{\cal D}_m$ denotes the 
superconformal covariant derivative. 
Here and hereafter we take the unit that 
the 5D Planck scale is unity, $M_5=1$. 

%\subsection{Lagrangian}
The invariant action for 5D supergravity on $S^1/Z_2$ 
is given as~\cite{Fujita:2001bd}  
\begin{eqnarray}
S &=& \int d^4x \int dy\, 
({\cal L}_b + {\cal L}_f + {\cal L}_{\rm aux} + {\cal L}_{N=1}), 
\nonumber
\end{eqnarray}
where ${\cal L}_b$, ${\cal L}_f$, ${\cal L}_{\rm aux}$ 
and ${\cal L}_{N=1}$ are the Lagrangians for the bosonic, 
fermionic, auxiliary fields and the boundary Lagrangian 
respectively, which are shown in Eqs.~(\ref{eq:bulklag}) and 
(\ref{eq:bdrylag}) in Appendix~\ref{app:trsact}. 
It was shown in Refs.~\cite{Fujita:2001bd,Kugo:2002js} that the above 
off-shell supergravity can be consistently compactified on an orbifold 
$S^1/Z_2$ by the $Z_2$-parity assignment shown in Table~\ref{tab:parity} 
in Appendix~\ref{app:trsact}.

\section{$SU(2)_{\mbox{\small \boldmath $U$}}$ 
gauge fixing and Scherk-Schwarz twist}
\label{sec:ssconf}
In this section, starting from the hypermultiplet compensator 
formalism reviewed in the previous section, we construct 
the 5D Poincar\'e supergravity with SS twist by introducing the 
twisted $SU(2)_{\mbox{\scriptsize \boldmath $U$}}$ gauge fixing. 
In the following subsections, 
we will explicitly show the Wilson line interpretation, 
the correspondence between the twist and the boundary constant 
superpotentials and the inconsistency of the twist on the 
warped geometry within our framework. 

% \subsection{Scherk-Schwarz basis}
First we recall the usual SS mechanism. 
The following arguments are based on the 5D Poincar\'e 
supergravity on the orbifold $S^1/Z_2$ defined by 
$$
\Phi(x,-y)= \pm Z\Phi(x,y), 
$$
where $Z$ acts on the $SU(2)_R$ indices of fields and 
is chosen as $Z=\sigma_3$ without a loss of generality. 
The SS twist is defined as 
$$
\Phi(x,y+2 \pi R)=T\Phi(x,y), 
$$
where $R$ is the radius of the fifth dimension $y$ and 
$T$ also acts on the $SU(2)_R$ indices that is given by 
$$
T=e^{-2\pi i \vec\omega \cdot \vec\sigma}. 
$$
The twist vector $\vec\omega = (\omega_1,\omega_2,\omega_3)$ 
determines the strength and the direction of the SS twist 
and $\vec\sigma = (\sigma_1,\sigma_2,\sigma_3)$ are the Pauli matrices. 
For $Z=\sigma_3$, the consistency condition
\begin{eqnarray}
TZT &=& Z, 
\label{eq:tzt}
\end{eqnarray}
requires $\omega_3=0$ 
(except for some special cases~\cite{vonGersdorff:2001ak}). 
We can go to the periodic field basis by redefining 
fields~\cite{vonGersdorff:2001ak,vonGersdorff:2003qf}
as $\Phi \to e^{i\vec\omega \cdot \vec\sigma\,f(y)} \Phi$ where 
$e^{i\vec\omega \cdot \vec\sigma\,f(y)}$ again acts on the 
$SU(2)_R$ indices and $f(y)$ satisfies 
\begin{eqnarray}
f(y+2 \pi R) &=& f(y) +2\pi. 
\label{eq:gfpa}
\end{eqnarray}

\subsection{Twisted $SU(2)_{\mbox{\small \boldmath $U$}}$ gauge fixing}
In the derivation of 5D Poincar\'e supergravity from the 
5D conformal supergravity using the hypermultiplet compensator, 
the $SU(2)_R$ symmetry is defined as the diagonal subgroup of 
the direct product between the original 
$SU(2)_{\mbox{\scriptsize \boldmath $U$}}$ gauge 
symmetry and $SU(2)_C$ which rotates the compensator index $a$, 
\begin{eqnarray}
SU(2)_{\mbox{\scriptsize \boldmath $U$}} \times SU(2)_C \to SU(2)_R, 
\label{eq:su2usu2r}
\end{eqnarray}
through the $\mbox{\boldmath $U$}$-gauge fixing 
${\cal A}^a_{\ i} \propto \delta^a_{\ i}$. 
The $\mbox{\boldmath $D$}$- and $\mbox{\boldmath $U$}$-gauge 
fixings in Eq.~(\ref{eq:scgf}) completely fix 
the quaternionic compensator hyperscalar field 
as\footnote{Note that the equations of motion for the auxiliary 
fields $D'$ and $\chi'$ in Eq.~(\ref{eq:bulklag}) in 
Appendix~\ref{app:trsact} result in ${\cal A}^2=2{\cal N}$ 
and ${\cal A}^{\bar\alpha}_{\ i} \zeta_\alpha
\equiv {\cal A}^a_{\ i} \zeta_a
-{\cal A}^{\underline\alpha}_{\ i} 
\zeta_{\underline\alpha}=0$ respectively.} 
\begin{eqnarray}
{\cal A}^a_{\ i} \equiv \delta^a_{\ i} 
\sqrt{1+{\cal A}^{\underline\alpha}_{\ i} 
{\cal A}_{\underline\alpha}^{\ i}}. 
\label{eq:norufix}
\end{eqnarray}
Then the compensator fermion $\zeta^a$ is fixed by 
\begin{eqnarray}
\zeta_a \delta^a_{\ i} 
&=& \zeta_{\underline\alpha} 
\frac{{\cal A}^{\underline\alpha}_{\ i}}
{\sqrt{1+{\cal A}^{\underline\beta}_{\ j} 
{\cal A}_{\underline\beta}^{\ j}}}. 
\label{eq:norsfix}
\end{eqnarray}

However, if we consider a torus compactification of the fifth 
dimension $y$ with the radius $R$, we have inequivalent 
classes of the $SU(2)_{\mbox{\scriptsize \boldmath $U$}}$ gauge fixing 
which is parameterized by a twist vector $\vec\omega$ as 
\begin{eqnarray}
{\cal A}^a_{\ i} \equiv 
\left( e^{i\vec\omega \cdot \vec\sigma\,f(y)} \right)^a_{\ i} 
\sqrt{1+{\cal A}^{\underline\alpha}_{\ i} {\cal A}_{\underline\alpha}^{\ i}}, 
\label{eq:twtufix}
\end{eqnarray}
and then 
\begin{eqnarray}
\zeta_a \left( e^{i\vec\omega \cdot \vec\sigma\,f(y)} 
\right)^a_{\ i}  
&=& \zeta_{\underline\alpha} 
\frac{{\cal A}^{\underline\alpha}_{\ i}}
{\sqrt{1+{\cal A}^{\underline\beta}_{\ j} 
{\cal A}_{\underline\beta}^{\ j}}}, 
\label{eq:twtsfix}
\end{eqnarray}
where $f(y)$ is a function satisfying Eq.~(\ref{eq:gfpa}). 
This twist vector $\vec\omega$ specifies the class of torus 
compactified 5D Poincar\'e supergravity and $\omega_{1,2} \ne 0$ 
corresponds to the SS twist parameter. In other words, 
from Eq.~(\ref{eq:su2usu2r}), the SS boundary condition 
for all the fields with $SU(2)_R$ index in the Poincar\'e 
supergravity is simply (equivalently) given by only the 
compensator ($SU(2)_C$) twisting in the framework 
of conformal supergravity~\cite{Abe:2004ar}. 
We usually choose $f(y)=y/R$, but here we do not specify its explicit 
form for the later convenience. Note that the different choice 
of $f(y)$ with the same $\vec\omega$ gives the same physics, 
because it corresponds to the 
$\mbox{\boldmath $U$}$-gauge fixing parameter. 

% \subsection{Periodic basis and Wilson line interpretation}
In the superconformal formulation, 
$SU(2)_{\mbox{\scriptsize \boldmath $U$}}$ 
gauge field on-shell (without the boundary action) is given by 
\begin{eqnarray}
V_{\mu\,{\rm sol}}^{ij} &=& 
-\frac{1}{2{\cal N}} \left( 
2{\cal A}^{\bar\alpha (i} \nabla_\mu {\cal A}_{\alpha}^{\ j)} 
+i{\cal N}_{IJ} \bar\Omega^{Ii} \gamma_\mu \Omega^{Jj} \right), 
\label{eq:onshellv}
\end{eqnarray}
where 
${\cal N}_{IJ}=\partial {\cal N}/\partial M^I \partial M^J$. 
The notations $(i$, $j)$ and $\bar\alpha$ in the superscript 
are defined respectively in Eqs.~(\ref{eq:symsum}) and 
(\ref{eq:barnotation}) in Appendix~\ref{app:trsact}. 
Under the gauge fixing (\ref{eq:twtufix}), 
we go to the periodic field basis by 
\begin{eqnarray}
{\cal A}^a_{\ i} \to U^a_{\ b}(y) {\cal A}^b_{\ i}, \qquad 
\zeta^a \to U^a_{\ b}(y) \zeta^b, 
\label{eq:bb}
\end{eqnarray}
where 
\begin{eqnarray}
U^a_{\ b}(y) \equiv 
\left( e^{-i\vec\omega \cdot \vec\sigma\,f(y)} \right)^a_{\ b}. 
\nonumber
\end{eqnarray}
In this basis all the field (including the compensator) 
has a usual periodic boundary condition. 
The $SU(2)_{\mbox{\scriptsize \boldmath $U$}}$ gauge fixing 
(\ref{eq:twtufix}) reduces into trivial one (\ref{eq:norufix}), 
while Eq.~(\ref{eq:onshellv}) becomes 
\begin{eqnarray}
{V}_{y\,{\rm sol}}^{ij} &\to& 
-\frac{1}{2{\cal N}} \left( 
2{\cal A}^{\bar\alpha (i} \nabla_y {\cal A}_{\alpha}^{\ j)} 
+ia_{IJ} \bar\Omega^{Ii} \gamma_y \Omega^{Jj} \right) 
+\frac{1}{\cal N}(U_{ca} \partial_y U^c_{\ b}) 
{\cal A}^{a(i} {\cal A}^{bj)}. 
\label{eq:onshellvhb}
\end{eqnarray}
Namely we find that the twisted 
$SU(2)_{\mbox{\scriptsize \boldmath $U$}}$ gauge fixing 
(\ref{eq:twtufix}) yields a field-independent shift 
of $SU(2)_{\mbox{\scriptsize \boldmath $U$}}$ gauge field $V_y^{ij}$. 
It leads to a nonvanishing contribution to the Wilson line, 
\begin{eqnarray}
\int_0^{2 \pi R} \!dy\, {V_y}^i_{\ j} 
&=& 2\pi \, i(\vec\omega \cdot \vec\sigma)^i_{\ j} + \cdots, 
\nonumber
\end{eqnarray}
where we have adopted the property (\ref{eq:gfpa}) and 
the ellipsis denotes the contributions from the physical fields. 
Then we find that the SS SUSY breaking 
can be interpreted as the breaking caused by a nonvanishing Wilson line 
of $SU(2)_{\mbox{\scriptsize \boldmath $U$}}$ gauge field. 
As shown in Ref.~\cite{vonGersdorff:2001ak} (based on the linear 
multiplet compensator formalism~\cite{Zucker:2000ks}), 
$V_y^{ij}$ appears in the auxiliary component of the 
so-called radion superfield in the $N=1$ superspace description. 
We will show this within our framework 
in Sec.~\ref{sec:ssss}. (See Eq.~(\ref{eq:radionmed}).) 

% \subsection{Breaking terms}
Next we derive SUSY breaking terms induced by the SS twist 
in the periodic basis. By the field redefinition (\ref{eq:bb}), 
the compensator fixing conditions become normal ones 
(\ref{eq:norufix}) and (\ref{eq:norsfix}) respectively, 
while we have additional $\vec\omega$ dependent terms which arise 
through the $y$-derivatives of the compensator fields in 
the action shown in Eq.~(\ref{eq:bulklag}) in Appendix~\ref{app:trsact}. 
They are given by 
\begin{eqnarray}
e^{-1}{\cal L}_\omega &=& 
f'(y) (i\vec\omega \cdot \vec\sigma)_{ab} 
\bigg\{ \epsilon^{ij} 
({\cal A}^a_{\ j}\nabla_4{\cal A}^b_{\ i} 
-{\cal A}^b_{\ i}\nabla_4{\cal A}^a_{\ j}) 
+2i\bar\zeta^b \gamma^4 \zeta^a 
\nonumber \\ &&
-4i\bar\psi_m^i \gamma^4 \gamma^m \zeta^a {\cal A}^b_{\ i}
+2i\bar\psi_m^{(i} \gamma^{m4n}\psi_n^{j)} 
{\cal A}^b_{\ j}{\cal A}^a_{\ i} 
-\frac{1}{\cal N}\epsilon^{jk}{V_4}^i_{\ k} 
({\cal A}^b_{\ i}{\cal A}^a_{\ j} 
+{\cal A}^b_{\ j}{\cal A}^a_{\ i}) \bigg\} 
\nonumber \\ &&
-(f'(y)|\vec\omega|)^2 \epsilon^{ij}\epsilon_{ab} 
{\cal A}^b_{\ i}{\cal A}^a_{\ j}. 
\nonumber
\end{eqnarray}
where $\epsilon^{ij}=\epsilon_{ij}=i\sigma_2$. 
We remark that the $\vec\omega$ dependent gaugino mass will 
be generated after integrating out the auxiliary field $V_4$. 

The $V_4$ is included in the auxiliary Lagrangian 
in Eq.~(\ref{eq:bulklag}) as 
$$e^{-1}{\cal L}_{\rm aux}^V=
-\eta^{mn} (V_m-V_{m\,{\rm sol}})^{ij} (V_n-V_{n\,{\rm sol}})_{ij},$$ 
where $V_{m\,{\rm sol}}=e^\mu_m V_{\mu\,{\rm sol}}$ 
is given by Eq.~(\ref{eq:onshellv}). Then ${\cal L}_\omega$ 
on-shell becomes 
\begin{eqnarray}
e^{-1}{\cal L}_\omega^{\rm on \textrm{-} shell} 
&=& 
% f'(y) (i\vec\omega \cdot \vec\sigma)_{ab} 
% \bigg\{ \epsilon^{ij} 
% ({\cal A}^a_{\ j}\nabla_4{\cal A}^b_{\ i} 
% -{\cal A}^b_{\ i}\nabla_4{\cal A}^a_{\ j}) 
% +2i\bar\zeta^b \gamma^4 \zeta^a 
% -4i\bar\psi_m^i \gamma^4 \gamma^m \zeta^a {\cal A}^b_{\ i}
% \nonumber \\ &&
% +2i\bar\psi_m^{(i} \gamma^{m4n}\psi_n^{j)} 
% {\cal A}^b_{\ j}{\cal A}^a_{\ i} 
% -\frac{1}{\cal N}\epsilon^{jk}{V_{4\,{\rm sol}}}^i_{\ k} 
% ({\cal A}^b_{\ i}{\cal A}^a_{\ j} 
% +{\cal A}^b_{\ j}{\cal A}^a_{\ i}) \bigg\} 
% \nonumber \\ &&
% -\frac{1}{4{\cal N}^2} (f'(y))^2 
% (i\vec\omega \cdot \vec\sigma)_{ab} 
% (i\vec\omega \cdot \vec\sigma)_{cd} 
% \epsilon^{ik} \epsilon^{jl} 
% ({\cal A}^b_{\ i}{\cal A}^a_{\ j}
% +{\cal A}^b_{\ j}{\cal A}^a_{\ i}) 
% ({\cal A}^d_{\ k}{\cal A}^c_{\ l}
% +{\cal A}^d_{\ l}{\cal A}^c_{\ k}) 
% \nonumber \\ &&
% -(f'(y)|\vec\omega|)^2 \epsilon^{ij}\epsilon_{ab} 
% {\cal A}^b_{\ i}{\cal A}^a_{\ j} 
% \nonumber \\ &=& 
f'(y) (i\vec\omega \cdot \vec\sigma)_{ij} \bigg\{ 
2i\bar\zeta^{\underline\beta} \gamma^4 \zeta^{\underline\alpha} 
{\cal A}_{\underline\beta}^{\ j} 
{\cal A}_{\underline\alpha}^{\ i} 
(1+{\cal A}_{\underline\alpha}^{\ k}
{\cal A}^{\underline\alpha}_{\ k})^{-1} 
+4i\bar\psi_m^i \gamma^4 \gamma^m 
\zeta^{\underline\alpha} {\cal A}_{\underline\alpha}^{\ j} 
\nonumber \\ &&
+\Big(2i\bar\psi_m^{(i} \gamma^{m4n}\psi_n^{j)}
-2{\cal A}^{\underline\alpha (i} \nabla_4 
{\cal A}_{\underline\alpha}^{\ j)} 
+ia_{IJ} \bar\Omega^{Ii} \gamma_4 \Omega^{Jj} \Big) 
(1+{\cal A}_{\underline\alpha}^{\ k}
{\cal A}^{\underline\alpha}_{\ k}) \bigg\} 
\nonumber \\ &&
-2(f'(y)|\vec\omega|)^2 
\big( {\cal A}_{\underline\alpha}^{\ i}
{\cal A}^{\underline\alpha}_{\ i}
+({\cal A}_{\underline\alpha}^{\ i}
{\cal A}^{\underline\alpha}_{\ i})^2 \big), 
\nonumber
\end{eqnarray}
where we have applied the compensator fixings 
(\ref{eq:norufix}) and (\ref{eq:norsfix}). 
Then the important part for the low energy physics 
up to quadratic in fields is given by 
\begin{eqnarray}
e^{-1}{\cal L}_\omega^{\rm mass} 
&=& f'(y) (i\vec\omega \cdot \vec\sigma)_{ij} \Big(
2i\bar\psi_m^{(i} \gamma^{m4n}\psi_n^{j)}
+ia_{IJ} \bar\Omega^{Ii} \gamma_4 \Omega^{Jj} \Big) 
-2(f'(y)|\vec\omega|)^2 
{\cal A}_{\underline\alpha}^{\ i}
{\cal A}^{\underline\alpha}_{\ i}, 
\nonumber
\end{eqnarray}
which contains the mass terms of the gravitino, the gauginos 
and the hyperscalars~\cite{Marti:2001iw,vonGersdorff:2001ak,
vonGersdorff:2003qf}. 
We remark that the cosmological constant proportional to 
$|\vec\omega|^2$ in ${\cal L}_\omega$ has been cancelled 
on-shell by the equation of motion for $V_4$. 

\subsection{Singular gauge fixing and boundary interpretation}
\label{sec:sscs}
As we mentioned above, an explicit function form of $f(y)$ 
does not affect the physical consequence because $f(y)$ is 
just a gauge fixing parameter. 
We usually choose $f(y)=y/R$ which gives the 
simplest description. 
However in this section, motivated by the argument of 
{\it generalized symmetry breaking} in Ref.~\cite{Bagger:2001qi}, 
we choose a gauge fixing parameter, 
\begin{eqnarray}
f(y) &=& \frac{\pi}{2} \sum_n 
\big( {\rm sgn}(y-n \pi R) - {\rm sgn}(-n \pi R) \big), 
\label{eq:singularufix} 
\end{eqnarray}
where ${\rm sgn}(y)$ is the sign-function. 
Namely, 
\begin{eqnarray*}
f'(y) &=& \pi \sum_n \delta(y-n \pi R). 
\end{eqnarray*}
For this $f(y)$, ${\cal L}_\omega$ becomes totally boundary terms. 
However we should be careful about the singular structure 
of this gauge fixing parameter which can change physics 
depending on the regularization. 
In the following we concentrate on the contribution 
from the $Z_2$-even fields 
which does not suffer from this singularity. 
We will briefly mention about the correction from 
the $Z_2$-odd fields at the end of this section. 

In the situation that the compensator is not charged 
under the physical gauge field (i.e. $R$-symmetry is 
not gauged), we introduce $N=1$ invariant constant 
superpotentials $W$ at the orbifold fixed points 
(see Eq.~(\ref{eq:bdrylag}) in Appendix~\ref{app:trsact}) 
\begin{eqnarray}
{\cal L}_{N=1} &=& 
\sum_n \delta(y-n\pi R) 
\left[ \Sigma^3 W \right]_F 
\nonumber \\ &=&
\sum_n \delta(y-n\pi R) e_{(4)} W 
\nonumber \\ && \times 
\bigg[ 2i{\cal A}^{a=2}_{\ i=2} 
(1+iW_4^{I=0}/M^{I=0}) \tilde{\cal F}^{a=2}_{\ i=1} 
+2{\cal A}^{a=2}_{\ i=2} 
\big( \nabla_4{\cal A}+{\cal A}V_4 
-2i\bar\psi_4 \zeta \big)^{a=2}_{\ i=1} 
\nonumber \\ &&
-2\bar\zeta_+ {\cal P}_R \zeta_+^{\hat\alpha=1} 
+4i{\cal A}^{a=2}_{\ i=2} 
\bar\psi_+ \cdot \gamma {\cal P}_R \zeta_+^{\hat\alpha=1} 
-2{\cal A}^{a=2}_{\ i=2} {\cal A}^{a=2}_{\ i=2} 
\bar\psi_{m+} \gamma^{mn} {\cal P}_L \psi_{n+} \bigg] 
+ {\rm h.c.}, \nonumber
\end{eqnarray}
where ${\cal P}_{R,L}\equiv (1 \pm \gamma_5)/2$, 
$\tilde{\cal F}^{\alpha}_{\ i}
\equiv {\cal F}^{\alpha}_{\ i}-{\cal F}^{\alpha}_{\ i\,{\rm sol}}$ 
and ${\cal F}^{\alpha}_{\ i\,{\rm sol}}$ is shown in 
Eq.~(\ref{eq:auxsol}) in Appendix~\ref{app:trsact}. 
The $Z_2$-even fields $\psi_{m+}$ and $\zeta_+^{\hat\alpha}$ 
are defined in Table~\ref{tab:parity} in Appendix~\ref{app:trsact}. 
The symbol $[\cdots]_F$ represents the $F$-term invariant formula 
in $N=1$ superconformal tensor calculus~\cite{Fujita:2001bd} 
and $e_{(4)}$ is the determinant of the 4D induced vierbein. 
Here let us see the relation between the SS twist 
and the boundary constant superpotentials. 
Interestingly enough, we can show that, 
provided $W$ satisfies 
\begin{eqnarray}
W=\pi (\omega_2+i\omega_1), 
\label{eq:sscsrel}
\end{eqnarray}
the constant superpotentials generate exactly the same terms 
as (the $Z_2$-even part of) the previous SS twist Lagrangian 
${\cal L}_\omega$ after integrating out 
compensator auxiliary field ${\cal F}^a_{\ i}$, 
that is, 
\begin{eqnarray}
{\cal L}_W +{\cal L}_{\rm aux}
&=& {\cal L}_\omega +{\cal L}_{\rm aux}
\big|_{{\cal F}^a_{\ i} \to {\cal F}^a_{\ i}+C^a_{\ i}}, 
\label{eq:sscsp}
\end{eqnarray}
where 
\begin{eqnarray}
C^a_{\ i} &=& 
\pi \sum_n \delta(y-n \pi R) e^{-1}e_{(4)} 
\left( \frac{(M^{I=0})^{-1} W_4^{I=0} \mathbf{1}_2 + {\sigma_3}}
{1+(M^{I=0})^{-2}(W_4^{I=0})^2} \right)^a_{\ b} 
(i\vec\omega \cdot \vec\sigma)^b_{\ c} {\cal A}^c_{\ i}. 
\nonumber
\end{eqnarray}
The Lagrangian  ${\cal L}_{\rm aux}$ is shown in 
Eq.~(\ref{eq:bulklag}) in Appendix~\ref{app:trsact} 
which consists of complete squares of the auxiliary fields 
including ${\cal F}^a_{\ i}$ and $V_y^{ij}$. 

Therefore we conclude that, with the singular 
gauge fixing parameter (\ref{eq:singularufix}), 
the $Z_2$-even part of the SS twist 
$\vec\omega=(\omega_1,\omega_2,0)$ is equivalent to the $N=1$ invariant 
constant superpotentials $W=\pi(\omega_2+i\omega_1)$ at boundaries. 
From this correspondence between SS twist and the constant 
superpotentials, we confirm that SUSY breaking caused by the SS 
twist is not explicit (at least for the $Z_2$-even part), because 
the boundary constant superpotential is $N=1$ invariant. 
We remark that this correspondence has been easily found 
at the full supergravity level, not in the effective theory, 
thanks to our simplified interpretation of SS twist. 

We should comment that with a suitable regularization of 
the $\delta$-function, the parity odd part on the boundary 
modifies effectively the relation (\ref{eq:sscsrel}) as, 
e.g. $W=\tan (\pi \omega_2)$ for the $\sigma_2$ twist 
$\vec\omega=(0,\omega_2,0)$, that was shown in 
Ref.~\cite{vonGersdorff:2001ak}.

\subsection{Scherk-Schwarz twist and AdS$_5$ geometry}
\label{sec:ssads5}
It was suggested in Ref.~\cite{Hall:2003yc} 
that the SS twist yields an inconsistency 
in the supergravity on AdS$_5$ geometry. 
Therefore, in this section, we consider the SS twist in the 
AdS$_5$ background within our framework of twisted 
$SU(2)_{\mbox{\scriptsize \boldmath $U$}}$ fixing, 
and see what happens. 
It is known that the gauging of $U(1)_R$ symmetry by the graviphoton 
is necessary to realize the AdS$_5$ geometry keeping 
SUSY~\cite{Altendorfer:2000rr,
Gherghetta:2000qt,Bergshoeff:2000zn}. 
In fact, the negative cosmological constant is proportional 
to the $U(1)_R$ gauge coupling in such a case. 
When a physical vector field $W_{R\mu}$ gauges the $U(1)_C$ 
subgroup of $SU(2)_C$, 
\begin{eqnarray}
{\cal D}_\mu {\cal A}^a_{\ i} &=& 
\partial_\mu {\cal A}^a_{\ i} - {\cal A}^{aj} V_{\mu ij} 
-(g_R W_\mu^R)^a_{\ b} {\cal A}^b_{\ i}+\cdots, 
\label{eq:covdercomp}
\end{eqnarray}
where 
\begin{eqnarray}
(g_R W_\mu^R)^a_{\ b} &\equiv& 
g_R W_\mu^R  (\vec{q} \cdot i\vec\sigma)^a_{\ b} 
\ = \ \left\{ 
\begin{array}{ll}
|\vec{q}|g_R W_\mu^R (i\sigma_1 \sin \theta_R 
+i\sigma_2 \cos \theta_R)^a_{\ b}
& (g_R:\textrm{$Z_2$-even}) \\
|\vec{q}|g_R W_\mu^R (i\sigma_3)^a_{\ b} & (g_R:\textrm{$Z_2$-odd}) 
\end{array} \right., 
\nonumber
\end{eqnarray}
we shift the auxiliary $SU(2)_{\mbox{\scriptsize \boldmath $U$}}$ 
gauge field by 
$V_{\mu ij}^N \equiv V_{\mu ij} + g_R W_{\mu ij}^R$~\cite{Kugo:2000af}. 
Then the covariant derivative (\ref{eq:covdercomp}) becomes 
\begin{eqnarray}
{\cal D}_\mu {\cal A}^a_{\ i} &=& 
\partial_\mu {\cal A}^a_{\ i} + {\cal A}^{aj} V_{\mu ij}^N 
-g_R {\cal A}^{aj} W_{\mu ij}^R 
-(g_R W_\mu^R)^a_{\ b} {\cal A}^b_{\ i}+\cdots. 
\label{eq:rgauging}
\end{eqnarray}
The third and fourth term in the right-hand side of 
Eq.~(\ref{eq:rgauging}) cancels each other under 
the normal $\mbox{\boldmath $U$}$-gauge fixing 
(\ref{eq:norufix}), and consequently this $W_\mu^R$ 
becomes the massless $U(1)_R$ gauge field. 

However, with the nonvanishing SS twist $\vec\omega$ in 
the twisted $\mbox{\boldmath $U$}$-gauge fixing (\ref{eq:twtufix}), 
we find that the compensator decouples from this $W_\mu^R$ 
vector field in (\ref{eq:rgauging}), when and only when 
the condition 
\begin{eqnarray}
\left[ (\vec{q} \cdot i\vec\sigma),\, 
(\vec{\omega} \cdot i\vec\sigma) \right] &=& 0, 
\label{eq:ccfrg}
\end{eqnarray}
is satisfied. 
Otherwise the $U(1)_R$ gauge field acquires a mass 
from the beginning (from the superconformal-covariant 
kinetic term of the compensator), 
which breaks the unitarity of this Poincar\'e supergravity. 
This means that the twist vector $\vec\omega$ should satisfy 
(\ref{eq:ccfrg}) in order to consistently gauge the $U(1)_R$ 
subgroup of $SU(2)_R$ symmetry defined by $\vec{q}$. 

To obtain the GP-FLP~\cite{Gherghetta:2000qt} 
(BKVP~\cite{Bergshoeff:2000zn}) model for a supersymmetric 
warped brane world, we need to gauge the $U(1)_R$ 
symmetry by the graviphoton with $Z_2$-odd gauge coupling\footnote{
The $Z_2$-odd gauge coupling can be realized in the supergravity 
through the four-form mechanism~\cite{Bergshoeff:2000zn}.}
$g_R$, i.e. $\vec{q}=|\vec{q}|(0,0,1)$.  From the above argument, 
the only possible twist vector in this case is $\vec\omega=0$, 
which does not cause SUSY breaking\footnote{Note that 
$\omega_3=0$ due to $TZT=Z$ in Eq.~(\ref{eq:tzt}).}. 

On the other hand in the ABN model~\cite{Altendorfer:2000rr} 
in which the $U(1)_R$ symmetry is gauged by the graviphoton with 
$Z_2$-even gauge coupling $g_R$, i.e. 
$\vec{q}=|\vec{q}|\,(\sin \theta_R,\cos \theta_R,0)$, 
the possible twist vector is 
$\vec\omega=|\vec\omega|\,(\sin \theta_R,\cos \theta_R,0)$. 
In this case we find a possibility to have the SS SUSY breaking 
in a slice of AdS$_5$. However it was pointed out in 
Refs.~\cite{Kugo:2002js,Zucker:2000ks} that ABN model is not 
derived from the known off-shell formulations with the linear 
multiplet~\cite{Zucker:2000ks} 
or the hypermultiplet compensator~\cite{Kugo:2002js}. 
We can not find the Killing spinor on this background 
in those off-shell formulations for ABN model 
even without the SS twisting. 
This is still an open question.

\section{$N=1$ superspace description}
\label{sec:ssss}
In this section, we describe SS SUSY breaking in the $N=1$ 
superspace formalism, which is directly derived from the 
5D conformal supergravity~\cite{PaccettiCorreia:2004ri,Abe:2004ar}. 
As we found in the previous section, the SS twist yields 
inconsistent theory when the $R$-symmetry is gauged by some 
physical gauge field. Then, in this section, we restrict ourselves 
to the case that the $R$-symmetry is not gauged, which means 
that the background geometry is flat (neglecting the backreaction 
from the VEVs of the physical fields). 
We first review the action shown in Ref.~\cite{Abe:2005ij}, 
and extend it to the case with twisted 
$SU(2)_{\mbox{\scriptsize \boldmath $U$}}$ 
gauge fixing (\ref{eq:twtufix}). 

For simplicity, we consider the case that $n_V=1$, $(p,q)=(1,1)$ 
and the maximally symmetric norm function, 
${\cal N}=(M^{I=0})^3-M^{I=0}(M^{I=1})^2/2$. 
Then, the superspace action is written by the Lagrangian 
${\cal L}={\cal L}_C+{\cal L}_{V+H}+{\cal L}_{N=1}$, 
where 
\begin{eqnarray}
{\cal L}_C &=& 
-2\int d^4 \theta\, V_E 
(\bar\Xi \Xi + \bar\Xi^c \Xi^c)
-\left\{ \int d^2 \theta\, 
(\Xi^c \partial_y \Xi - \Xi \partial_y \Xi^c) 
+\textrm{h.c.} \right\}, 
\label{eq:ssss} \\
{\cal L}_{V+H} &=& 
\left\{ \int d^2\theta\,\frac{1}{4} 
E {\cal W} {\cal W} +\textrm{h.c.} \right\} 
+\int d^4 \theta\, 
\frac{E+\bar{E}}{2V_E^2}\left( -\partial_y V 
-i\Phi_S+i\bar\Phi_S \right)^2 
\nonumber \\ &&
+\int d^4 \theta\, V_E 
\left( \bar{H} e^{2gV} H 
+\bar{H}^c e^{-2gV} H^c \right) 
\nonumber \\ &&
+\left\{ 
\int d^2 \theta\, H^c 
\left( \partial_y 
+m\,E -2ig \Phi_S \right) H +\textrm{h.c.} 
\right\}, 
\nonumber
\end{eqnarray}
and ${\cal L}_{N=1}$ is the boundary Lagrangian 
for which we omit the explicit expressions. 
The superfields ($\Xi$, $\Xi^c$) and ($H$, $H^c$) represent 
the $N=1$ chiral multiplets which originate 
from the compensator and physical hypermultiplets respectively, 
and the superfields ($V$, $\Phi_S$) are the $N=1$ vector and 
chiral multiplets coming from the physical 5D vector multiplet. 
The remaining superfields $V_E$ and $E$ are a spurion-like 
vector superfield coming from the fifth component of the 5D Weyl multiplet 
and a chiral superfield coming from the central charge vector multiplet, 
respectively. 
Especially $E$ corresponds to the radion superfield in the 
`5D off-shell approach' in the terminology of Ref.~\cite{Abe:2005ij}. 
Note that $V_E \simeq (E+\bar{E})/2$ for the vanishing 
VEVs of physical fields. 
The mass scales are again measured in the unit of $M_5 \equiv 1$. 
Note that we have omitted the terms involving graviphoton ($I=0$) 
multiplet and a part of Chern-Simons terms, 
that are irrelevant to the following discussions. 

The relation between the fields in the 
superconformal multiplets and the above superfields 
$\Xi$, $\Xi^c$, $H$, $H^c$, $V$, $\Phi_S$, $V_E$ and $E$ 
is found in the appendix of Ref.~\cite{Abe:2005ij}. 
The (untwisted) superconformal gauge fixings (\ref{eq:scgf}) 
for the compensator chiral superfields are rewritten as 
\begin{eqnarray}
\xi &=& 
\sqrt{1+\frac{1}{2}(|h|^2+|h^c|^2)}, 
\qquad 
\xi^c \ = \ 0, 
\nonumber \\
\chi_\xi &=& 
\frac{1}{2\xi}(\bar{h}\chi_h+\bar{h}^c\chi_h^c), 
\qquad 
\chi_\xi^c \ = \ 
-\frac{1}{2\xi}(h\chi_h^c+h^c\chi_h), 
\label{eq:xifix}
\end{eqnarray}
where each component field is defined as 
$\Phi=\phi-\theta \chi_\phi-\theta^2 {\cal F}_\phi$, 
$\Phi^c=\phi^c-\theta \chi_\phi^c-\theta^2 {\cal F}_\phi^c$ 
($\Phi=\Xi,H$). 

Now we consider the twisted $SU(2)_{\mbox{\scriptsize \boldmath $U$}}$-gauge fixing 
(\ref{eq:twtufix}), 
which corresponds to the SS-twisted 
Poincar\'e supergravity. In this case, thanks to our simplified 
interpretation, the only change is the replacement of the 
compensator chiral superfields $\Xi$ and $\Xi^c$ by 
\begin{eqnarray}
\left( 
\begin{array}{c}
\Xi^c \\ \Xi 
\end{array}
\right) 
&\to& 
e^{-i\vec\omega \cdot \vec\sigma f(y)} 
\left( 
\begin{array}{c}
\Xi^c \\ \Xi 
\end{array}
\right). 
\label{eq:twistedcsf}
\end{eqnarray}
This replacement just generates additional terms in the action, 
\begin{eqnarray}
{\cal L}_{\rm twist} &=& 
f'(y) \left\{ 
(\omega_2+i\omega_1) \int d^2 \theta\, \Xi^2 
+(\omega_2-i\omega_1) \int d^2 \theta\, (\Xi^c)^2 
+\textrm{h.c.} \right\}, 
\label{eq:twistsslag}
\end{eqnarray}
which originate from the $\partial_y \Xi$ and $\partial_y \Xi^c$ 
terms in Eq.~(\ref{eq:ssss}). 

We have to go to the on-shell description here, 
because $\Xi$, $\Xi^c$, and $E$ (as well as the 
graviphoton multiplet omitted here) become {\it dependent} 
superfields after the superconformal gauge fixings~\cite{Abe:2005ij} 
(see Eq.~(\ref{eq:xifix}) for $\Xi$ and $\Xi^c$). 
However in the global SUSY limit ($M_5 \to \infty$), 
these dependent superfields are reduced to spurion-like superfields, 
\begin{eqnarray}
\Xi &\simeq& 1-\theta^2 {\cal F}_\xi, 
\qquad 
\Xi^c \ \simeq \ -\theta^2 {\cal F}_\xi^c, 
\qquad 
E \ \simeq \ 1-\theta^2 {\cal F}_E, 
\label{eq:xispurion}
\end{eqnarray} 
and we can formally keep the superspace structure. 
From Eq.~(\ref{eq:xispurion}) together with the additional 
Lagrangian term (\ref{eq:twistsslag}), we find 
\begin{eqnarray}
{\cal L}_C &=& 
-2(|{\cal F}_\Xi|^2+|{\cal F}_\Xi^c|^2)
-\bar{\cal F}_E {\cal F}_\Xi -\bar{\cal F}_\Xi {\cal F}_E 
+(\partial_y {\cal F}_\Xi^c+\textrm{h.c.}) 
\nonumber \\ &&
+2f'(y) \left\{ (\omega_2+i\omega_1){\cal F}_\Xi 
+\textrm{h.c.} \right\} 
\nonumber \\ 
&=& -2(|{\cal F}_\Xi|^2+|{\cal F}_\Xi^c|^2)
-\bar{F}_E {\cal F}_\Xi 
-\bar{\cal F}_\Xi F_E 
+(\partial_y {\cal F}_\Xi^c+\textrm{h.c.}), 
\nonumber
\end{eqnarray}
where 
\begin{eqnarray}
F_E \equiv {\cal F}_E -2f'(y)(\omega_2-i\omega_1). 
\label{eq:radionmed}
\end{eqnarray}
Then we conclude that the effect of the SS twist 
is completely absorbed by the constant shift of ${\cal F}_E$ 
as expected. As discussed in the previous 
works~\cite{Marti:2001iw} 
we can easily obtain the soft SUSY breaking terms for 
the matter fields in ${\cal L}_{V+H}$ by substituting 
${\cal F}_E=F_E+2f'(y)(\omega_2-i\omega_1)$ 
in the $F$-component of the spurion-like superfield $E$. 
Note that all the terms generated by integrating out 
$F_E$, ${\cal F}_\Xi$ and ${\cal F}_\Xi^c$ are 
higher order in powers of $1/M_5$, and the 
leading order effects of the SS SUSY 
breaking are obtained just by substituting 
${\cal F}_E=2f'(y)(\omega_2-i\omega_1)$ 
in ${\cal L}_{V+H}$. 

Finally we should remark that the correspondence between 
the SS twist and the constant superpotentials 
discussed in the previous section is manifest in this 
superspace description, that is, Eq.~(\ref{eq:twistsslag}) is 
nothing but the $N=1$ invariant constant superpotential. 
%by noticing that the Weyl weight of $\Xi$ is $3/2$. 
For the case of the singular gauge fixing (\ref{eq:singularufix}), 
this is the boundary constant superpotential, 
as shown in the previous section at the full 
superconformal gravity level. 
(See the relation (\ref{eq:sscsrel}).)

\section{Conclusions and discussions}
\label{sec:conclusion}
By noticing that the twisted $SU(2)_R$ boundary condition 
in $S^1$-compactified 5D Poincar\'e supergravity is equivalent 
to the twisted $SU(2)_{\mbox{\scriptsize \boldmath $U$}}$ 
gauge fixing in 5D conformal supergravity, we have reexamined 
the SS mechanism of SUSY breaking in the latter terminology. 
In this case, only the compensator hypermultiplet is relevant 
to the SS breaking, and various properties of the SS mechanism 
can be easily understood. We realized the 5D Poincar\'e 
supergravity with the SS twist starting from the hypermultiplet 
compensator formalism of 5D conformal 
supergravity~\cite{Kugo:2000af,Fujita:2001kv,Fujita:2001bd,Kugo:2002js}. 
Thanks to this simplified interpretation, we can explicitly show 
the Wilson line interpretation of the SS twist~\cite{vonGersdorff:2001ak}, 
the correspondence between the twist and 
the constant superpotentials~\cite{Bagger:2001qi}, 
and the quantum inconsistency of the 
twist in the AdS$_5$ background~\cite{Hall:2003yc} 
at the full supergravity level. 
We have also derived the $N=1$ superspace 
description of the SS twist based on our previous 
works~\cite{Abe:2004ar,Abe:2005ac,Abe:2005ij}. 

We remark that the $N=1$ superspace description 
of SS twist is possible only in the `5D off-shell approach' 
in the terminology of Ref.~\cite{Abe:2005ij}, 
in which the SS twist is translated into the nonvanishing $F$-term of 
the radion superfield~$E$. 
When we use the word ``the radion superfield'', 
we have to specify the context in which we work. 
%In `4D off-shell approach' of Ref.~\cite{Abe:2005ij}, 
%the radion superfield can be treated as dynamical one. 
For example, we have the radion superfield also 
in the `4D off-shell approach' of Ref.~\cite{Abe:2005ij}, 
but the nonvanishing $F$-term of this radion superfield does not 
realize the SS twist. 
This is because this approach is possible only when the 
background preserves $N=1$ SUSY. For the SUSY breaking 
case such as the SS-twisted supergravity, we are forced 
to work in the 5D off-shell approach, in which the 
radion superfield is a spurion-like superfield. 

As shown in Sec.~\ref{sec:ssads5}, if the compensator 
is charged under some gauge field including the graviphoton 
(i.e. the $R$-symmetry is gauged), the SS twist generates a mass 
for the corresponding gauge field without the Higgs mechanism, 
except for the case of ABN type gauging~\cite{Altendorfer:2000rr}. 
This is the reason why the SS twist results in an inconsistent 
theory in the AdS$_5$ background~\cite{Hall:2003yc}. 
However, this does not mean that the constant superpotentials 
cannot be introduced at the fixed points of orbifold 
in the AdS$_5$ background, because the correspondence 
between the SS twist and the constant superpotentials shown in 
Sec.~\ref{sec:sscs} is only valid {\it without} the $R$-gaugings. 
Indeed the $R$-gauge field does not receive a mass from 
the constant superpotential. 
In such a sense, the constant superpotential belongs to 
a much wider class of deformation parameter than the SS twist 
in 5D supergravity on $S^1/Z_2$ orbifold. 

Then it may be interesting if we consider the boundary 
constant superpotentials with the $R$-symmetry gauged by 
the $Z_2$-odd graviphoton through the $Z_2$-odd gauge coupling 
(GP-FLP type gauging~\cite{Gherghetta:2000qt}), which can not 
be resembled by SS twist. This possibility is studied in the 
global SUSY limit in the first paper of Ref.~\cite{Marti:2001iw}. 
If the $R$-gauge coupling is $Z_2$-even (ABN type gauging), 
the SS twist is also allowed as shown in Sec.~\ref{sec:ssads5}, 
and the combination of the twist and the constant superpotentials 
may result in some nontrivial on-shell supergravities. 
We expect that this case may be related to the on-shell models 
with detuned brane tensions~\cite{Bagger:2002rw}, even though 
we have still some gap between the ABN model and the 
off-shell formulations as mentioned at the end of 
Sec.~\ref{sec:sscs}.  
On the other hand, for the $R$-symmetry gauged by the $Z_2$-even 
physical vector field~\cite{Abe:2004nx}, we can not introduce 
any constant superpotential because the superpotential must 
have nonvanishing $R$-charge at boundaries in this case. 

One of the advantages of our derivation of SS-twisted 
Poincar\'e supergravity is that, in addition to its simplicity, 
it is based on the superconformal formulation which can 
interpolate various frames (such as the so-called Einstein 
frame or string frame), or even various on-shell supergravities 
that are physically different from each other. 
It is well known that the superconformal formulation is indeed 
powerful in four dimensions. 
Our method is expected to be useful if we treat the 
SS-twisted 5D Poincar\'e supergravity as an effective theory 
of further higher-dimensional theory such as, e.g. 
the eleven-dimensional supergravity. 
For the eleven-dimensional supergravity compactified on 
$S^1/Z_2$ times Calabi-Yau three-fold, it is known that 
the effective 5D supergravity is derived by introducing 
two compensator hypermultiplets~\cite{Fujita:2001bd} 
in 5D conformal supergravity. 
This will be one of the extension of this paper 
in future works.

\subsection*{Acknowledgement}
The authors would like to thank Kiwoon~Choi 
for discussions at the early stage of this work. 
Y.S. was supported by the Japan Society for the 
Promotion of Science for Young Scientists (No.0509241).

\appendix
\section{Invariant action in hypermultiplet compensator formalism}
\label{app:trsact}
In this appendix we review the invariant action derived in 
Refs.~\cite{Kugo:2000af,Fujita:2001kv,Fujita:2001bd,Kugo:2002js} 
for the relevant multiplets to this paper. 

The action for 5D supergravity on $S^1/Z_2$ orbifold 
is given by~\cite{Fujita:2001bd}  
\begin{eqnarray}
S &=& \int d^4x \int dy\, 
({\cal L}_b + {\cal L}_f + {\cal L}_{\rm aux} + {\cal L}_{N=1}), 
\nonumber
\end{eqnarray}
where ${\cal L}_b$, ${\cal L}_f$, ${\cal L}_{\rm aux}$ 
and ${\cal L}_{N=1}$ are the Lagrangians for the bosonic, 
fermionic, auxiliary fields and the boundary Lagrangian, 
respectively given by 
\begin{eqnarray}
e^{-1}{\cal L}_b &=& 
-\frac{1}{2}{\cal N}R 
- \frac{1}{4}{\cal N}a_{IJ}F^I_{\mu \nu} F^{\mu \nu J} 
+ \frac{1}{2}{\cal N}a_{IJ} \nabla^m M^I \nabla_m M^J 
+\frac{1}{2} {\cal N}^{IJ} {\cal Y}^{ij}_I {\cal Y}_{Jij}
\nonumber \\ && 
+ \nabla^m {\cal A}^{\bar\alpha}_{\ i} \nabla_m {\cal A}_\alpha^{\ i} 
+ {\cal A}^{\bar\alpha}_{\ i} 
(g^2 M^2)_{\alpha}^{\ \beta} {\cal A}_{\beta}^{\ i} 
+ {\cal N}^{-1}({\cal A}^{\bar\alpha i} \nabla_a {\cal A}_\alpha^{\ j})^2 
\nonumber \\ && 
+ \frac{1}{8} e^{-1} 
c_{IJK} \epsilon^{\lambda \mu \nu \rho \sigma} W^I_\lambda 
\left( F^J_{\mu \nu} F^K_{\rho \sigma}+\frac{1}{2}g[W_\mu, W_\nu]^J
F^K_{\rho \sigma}  +\frac{1}{10}g^2 [W_\mu, W_\nu]^J 
[W_\rho, W_\sigma]^K \right), 
\nonumber \\*[10pt]
% \end{eqnarray}
% \begin{eqnarray}
e^{-1}{\cal L}_f &=& 
-2i{\cal N} \bar\psi_\mu \gamma^{\mu \nu \rho} \nabla_\nu \psi_\rho 
+2i{\cal N} a_{IJ} \bar\Omega^I \nabla \!\!\!\!/\, \Omega^J 
-2i \bar\zeta^{\bar\alpha} (\nabla \!\!\!\!/\, + gM) \zeta_\alpha 
\nonumber \\ &&
-2i {\cal A}^{\bar\alpha}_{\ i} \nabla_n {\cal A}_{\alpha j} 
\bar\psi_m^{\ (i} \gamma^{mnk} \psi_k^{\ j)} 
+2ia_{IJ} {\cal A}^{\bar\alpha}_{\ i} \nabla_a {\cal A}_{\alpha j} 
\bar\Omega^I_i \gamma^a \Omega^J_j 
-4i \nabla_n {\cal A}^{\bar\alpha}_{\ i} 
\bar\psi_m^i \gamma^n \gamma^m \zeta_\alpha 
\nonumber \\ &&
+{\cal N}(\bar\psi_m \psi_n) 
(\bar\psi_k \gamma^{mnkl} \psi_l + \bar\psi^m \psi^n) 
+ {\cal N}(a_{IJ} \bar\Omega^{Ii} \gamma_a \Omega^{Jj})^2 
-ig{\cal N}_I [\bar\Omega,\Omega]^I 
\nonumber \\ &&
+\frac{1}{8} (\bar\psi_k \gamma^{mnkl} \psi_l 
+ 2 \bar\psi^m \psi^n + a_{JK} \bar\Omega^J \gamma^{mn} \Omega^K 
+ \bar\zeta^{\bar\alpha} \gamma^{mn} \zeta_\alpha)^2 
\nonumber \\ &&
+\frac{i}{4} {\cal N}_I F_{mn}^I (W) 
(\bar\psi_k \gamma^{mnkl} \psi_l + 2 \bar\psi^m \psi^n 
+ a_{JK} \bar\Omega^J \gamma^{mn} \Omega^K 
+ \bar\zeta^{\bar\alpha} \gamma^{mn} \zeta_\alpha) 
\nonumber \\ &&
+i{\cal N}a_{IJ} \bar\psi_m 
(\gamma \cdot F^I(W) -2 \nabla \!\!\!\!/\, M^I) \gamma^m \Omega^J 
\nonumber \\ &&
-2{\cal N}a_{IJ} (\bar\Omega^I \gamma^m \gamma^{nk} \psi_m) 
(\bar\psi_n \gamma_k \Omega^J) 
+2{\cal N}a_{IJ} (\bar\Omega^I \gamma^m \gamma^n \psi_m) 
(\bar\psi_n \Omega^J) 
\nonumber \\ &&
+\frac{1}{4}i{\cal N}_{IJK} \bar\Omega^I \gamma \cdot F^J(W) \Omega^K 
-\frac{2}{3} (\bar\Omega^I \gamma^{mn} \Omega^J \bar\psi_m \gamma_n \Omega^K 
+\bar\psi^i \cdot \gamma \Omega^{Ij} \bar\Omega^J_{(i} \Omega^K_{j)}), 
\nonumber \\*[10pt]
% \end{eqnarray}
% \begin{eqnarray}
e^{-1} {\cal L}_{\rm aux} &=& 
D'({\cal A}^2+2{\cal N}) 
-8i \bar\chi^{'i} {\cal A}^{\bar\alpha}_{\ i} \zeta_\alpha 
-\frac{1}{2}{\cal N}_{IJ}
(Y^{Iij}-Y^{Iij}_{{\rm sol}})(Y^J_{ij}-Y^J_{ij\,{\rm sol}}) 
\nonumber \\ &&
+2{\cal N}(v-v_{\rm sol})^{mn}(v-v_{\rm sol})_{mn} 
-{\cal N}(V_\mu-V_{{\rm sol}\,\mu})^{ij}(V^\mu-V_{{\rm sol}}^\mu)_{ij} 
\nonumber \\ &&
+\big( 1-W^{I=0}_\mu W^{I=0\,\mu}/(M^{I=0})^2 \big) 
({\cal F}^{\bar\alpha}_{\ i}-{\cal F}^{\bar\alpha}_{\ i\,{\rm sol}}) 
({\cal F}_\alpha^{\ i}-{\cal F}_{\alpha\ {\rm sol}}^{\ i}), 
\label{eq:bulklag}
\end{eqnarray}
and 
\begin{eqnarray}
{\cal L}_{N=1} 
&=& \sum_{l=0,\pi} \Lambda_l \delta(y-y_l)
\Big( {\textstyle -\frac{3}{2}}
\big[ \Sigma \bar\Sigma e^{-K^{(l)}(S,\bar{S})/3} \big]_D
\nonumber \\ && \qquad 
+ \big[ f^{(l)}_{IJ} (S) W^{I \alpha} W^J_\alpha \big]_F
+ \big[ \Sigma^3 W^{(l)}(S) \big]_F \Big). 
\label{eq:bdrylag}
\end{eqnarray}
Here we have used some expressions defined as 
\begin{eqnarray}
{\cal F}^{\alpha}_{\ i\,{\rm sol}} 
&=& -M^{I=0}(gt_{I=0})^\alpha_{\ \beta}{\cal A}^\beta_{\ i}, 
\nonumber \\
V_{{\rm sol}\,m}^{ij} &=& 
-\frac{1}{2{\cal N}} \left( 
2 {\cal A}^{\bar\alpha (i} \nabla_m {\cal A}_\alpha^{\ j)} 
-i{\cal N}_{IJ} \bar\Omega^{Ii} \gamma_m \Omega^{Jj} \right), 
\nonumber \\
v_{{\rm sol}\,mn} &=& 
-\frac{1}{4{\cal N}} \left\{ 
{\cal N}_I F_{mn}^I(W) -i \left( 
6{\cal N} \bar\psi_m \psi_n 
+ \bar\zeta^{\bar\alpha} \gamma_{mn} \zeta_\alpha 
-\frac{1}{2} {\cal N}_{IJ} \bar\Omega^I \gamma_{mn} \Omega^J 
\right) \right\}, 
\nonumber \\
Y^{Iij}_{{\rm sol}} &=& 
{\cal N}^{IJ} {\cal Y}_J^{ij}, \qquad 
{\cal F}^\alpha_{{\rm sol}\,i} \ = \ 
M^{I=0}(gt_{I=0})^\alpha_{\ \beta} {\cal A}^\beta_{\ i}, 
\nonumber \\
{\cal Y}_I^{ij} &=& 
2 {\cal A}_\alpha^{\ (i} (gt_I)^{\bar\alpha \beta} {\cal A}_\beta^{\ j)} 
+i {\cal N}_{IJK} \bar\Omega^{Ji} \Omega^{Kj}, 
\label{eq:auxsol}
\end{eqnarray}
and $D'$, ${{\chi^i}'}$ are shifted auxiliary fields 
from $D$, $\chi^i$ in the Weyl multiplet respectively 
for which we omit the explicit form (see Ref.~\cite{Fujita:2001bd}). 
The notation $A^{(i}B^{j)}$ is defined as 
\begin{eqnarray}
A^{(i}B^{j)}=\frac{1}{2}(A^iB^j+A^jB^i). 
\label{eq:symsum}
\end{eqnarray}

The indices $I,J,\ldots = 0,1,2,\ldots,n_V$ label the 
vector multiplets, and $I=0$ component corresponds to 
the central charge vector (almost graviphoton) multiplet. 
The gauge kinetic mixing $a_{IJ}$ is calculated as 
\begin{eqnarray}
a_{IJ} &\equiv& 
-\frac{1}{2} \frac{\partial^2}{\partial M^I \partial M^J} \ln {\cal N}
\ = \ -\frac{1}{2{\cal N}} 
\left( {\cal N}_{IJ} - \frac{{\cal N}_I {\cal N}_J}{{\cal N}} \right), 
\nonumber
\end{eqnarray}
where ${\cal N}$ is the norm function of 5D supergravity 
explained in Sec.~\ref{sec:review}. 
The $n_V+1$ gauge scalar fields $M^I$ are constrained by 
$\mbox{\boldmath $D$}$ gauge fixing shown in Eq.~(\ref{eq:scgf}) 
resulting $n_V$ independent degrees of freedom. 

\begin{table}[t]
\begin{center}
\begin{tabular}{c|c}
\hline 
\multicolumn{2}{l}{\hfill 
\begin{minipage}{0.18\linewidth}
\ \\*[-2pt]
Weyl multiplet \\*[-9pt]
\end{minipage}
\hfill \hspace{0pt}} \\ 
\hline 
$\Pi=+1$ & 
\begin{minipage}{0.65\linewidth}
\ \\*[0pt]
\centerline{
$e_{\underline\mu}^{\ \underline{m}},\,e_y^{\ 4},\,
\psi_{\underline\mu+},\,\psi_{y-},\,\varepsilon_+,\,\eta_-,\,
b_{\underline\mu},\,V_{\underline\mu}^{(3)},\,V_y^{(1,2)},\,
v^{4\underline{m}},\,\chi_+,\,D$} \\*[-5pt]
\end{minipage} 
\\ \hline 
$\Pi=-1$ & 
\begin{minipage}{0.65\linewidth}
\ \\*[0pt]
\centerline{
$e_{\underline\mu}^{\ 4},\,e_y^{\ \underline{m}},\,
\psi_{\underline\mu-},\,\psi_{y+},\,\varepsilon_-,\,\eta_+,\,
b_y,\,V_y^{(3)},\,V_{\underline\mu}^{(1,2)},\,
v^{\underline{m}\underline{n}},\,\chi_-$} \\*[-5pt]
\end{minipage}
\\ \hline \hline 
\multicolumn{2}{l}{\hfill 
\begin{minipage}{0.2\linewidth}
\ \\*[-2pt]
Vector multiplet \\*[-9pt]
\end{minipage}
\hfill \hspace{0pt}} \\ 
\hline 
$\Pi(M)$ & 
\begin{minipage}{0.65\linewidth}
\ \\*[0pt]
\centerline{
$M,\,W_y,\,Y^{(1,2)},\,\Omega_-$} \\*[-6pt]
\end{minipage}
\\ \hline 
$-\Pi(M)$ & 
\begin{minipage}{0.65\linewidth}
\ \\*[0pt]
\centerline{
$W_{\underline\mu},\,Y^{(3)},\,\Omega_+$} \\*[-6pt]
\end{minipage} 
\\ \hline \hline 
\multicolumn{2}{l}{\hfill 
\begin{minipage}{0.18\linewidth}
\ \\*[-2pt]
Hypermultiplet \\*[-9pt]
\end{minipage}
\hfill \hspace{0pt}} 
\\ \hline 
$\Pi({\cal A}^{2\hat\alpha-1}_{\ i=1})$ & 
\begin{minipage}{0.65\linewidth}
\ \\*[0pt]
\centerline{
${\cal A}^{2\hat\alpha-1}_{\ i=1},\,
{\cal A}^{2\hat\alpha}_{\ i=2},\,
{\cal F}^{2\hat\alpha-1}_{\ i=2},\,
{\cal F}^{2\hat\alpha}_{\ i=1},\,
\zeta^{\hat\alpha}_+$} \\*[-6pt]
\end{minipage}
\\ \hline 
$-\Pi({\cal A}^{2\hat\alpha-1}_{\ i=1})$ & 
\begin{minipage}{0.65\linewidth}
\ \\*[0pt]
\centerline{
${\cal A}^{2\hat\alpha-1}_{\ i=2},\,
{\cal A}^{2\hat\alpha}_{\ i=1},\,
{\cal F}^{2\hat\alpha-1}_{\ i=1},\,
{\cal F}^{2\hat\alpha}_{\ i=2},\,
\zeta^{\hat\alpha}_-$} \\*[-6pt]
\end{minipage}
\\ \hline 
% \hline
% \multicolumn{2}{l}{\hfill Linear multiplet \hfill \hspace{0pt}} \\ \hline
% $\Pi(L^{(1,2)})$ & 
% $L^{(1,2)},\,N,\,E^4,\,\varphi_+$ \\ \hline
% $-\Pi(L^{(1,2)})$ & 
% $L^{(3)},\,E^{\underline{m}},\,\varphi_-$ \\ \hline
\end{tabular}
\end{center}
\caption{The $Z_2$ parity assignment. 
The underbar means that the index is 4D one. 
The subscript $\pm$ of the $SU(2)$ Majorana spinors 
is defined as, e.g. 
$\psi_+=\psi^{i=1}_R+\psi^{i=2}_L$ and 
$\psi_-=i(\psi^{i=1}_L+\psi^{i=2}_R)$ where 
$\psi_{R,L}=(1 \pm \gamma_5)\psi/2$, except for 
$\zeta_+^{\hat\alpha}=i(\psi^{\alpha=2\hat\alpha-1}_L
+\psi^{\alpha=2\hat\alpha}_R)$ and 
$\zeta_-^{\hat\alpha}=\psi^{\alpha=2\hat\alpha-1}_R
+\psi^{\alpha=2\hat\alpha}_L$ 
($\hat\alpha=1,\ldots,p+q$).}
\label{tab:parity}
\end{table}

For hyperscalars ${\cal A}^\alpha_{\ i}$, the index $i=1,2$ is the 
$SU(2)_{\mbox{\scriptsize \boldmath $U$}}$-doublet index and 
$\alpha=1,2,\ldots,2(p+q)$ where $p$ and $q$ are the numbers of the 
quaternionic compensator and physical hypermultiplets respectively. 
The notation $\bar\alpha$ in the action stands for 
\begin{eqnarray}
{\cal A}^{\bar\alpha}_{\ i} &=& 
d_{\beta}^{\ \alpha} {\cal A}^\beta_{\ i}, 
\label{eq:barnotation}
\end{eqnarray}
where 
$d_{\alpha}^{\ \beta} \equiv {\rm diag}(\mathbf{1}_{2p},-\mathbf{1}_{2q})$. 

It was shown in Ref.~\cite{Fujita:2001bd,Kugo:2002js} that the above 
off-shell supergravity can be consistently compactified on an orbifold 
$S^1/Z_2$ by the $Z_2$-parity assignment shown in Table~\ref{tab:parity} 
without a loss of generality. 
When the fifth dimension is compactified on such orbifold $S^1/Z_2$, 
we generically have the boundary $N=1$ invariant action 
${\cal L}_{N=1}$ with the constant $\Lambda_{0,\pi}$ at the orbifold 
fixed points $(y_0,y_\pi)=(0,\pi R)$ as shown in Eq.~(\ref{eq:bdrylag}). 
In the boundary action, $\Sigma$ is the 4D $N=1$ compensator chiral 
multiplet with the Weyl and chiral weight $(w,n)=(1,1)$ induced by 
the 5D compensator hypermultiplet, while $S$ and $W^{I\alpha}$ stand for 
generic chiral matter and gauge (field strength) multiplets 
with $(w,n)=(0,0)$ at the boundaries which come from either bulk fields 
or pure boundary fields. 
Here the symbols $[\cdots]_D$ and $[\cdots]_F$ represent the $D$- and 
$F$-term invariant formulae, respectively, in the $N=1$ superconformal 
tensor calculus~\cite{Fujita:2001bd}.

{\it Without} the boundary $N=1$ action 
${\cal L}_{N=1}$, the auxiliary fields on-shell are given by 
$V_\mu=V_{{\rm sol}\,\mu}$, 
$v_{mn}=v_{{\rm sol}\,mn}$, 
$Y^{Iij}=Y^{Iij}_{{\rm sol}}$, 
${\cal F}^\alpha_{\ i}={\cal F}^\alpha_{{\rm sol}\,i}$, 
and ${\cal L}_{\rm aux}$ finally vanishes on-shell, 
$e^{-1}{\cal L}_{\rm aux}^{\rm on \textrm{-} shell}=0$.

\end{document}